\documentclass[a4paper,11pt]{article}
\usepackage{pos}
\usepackage{graphicx}
\usepackage{subcaption}

\newcommand{\DKL}{D_{\mathrm{KL}}}

\newcommand{\dd}{\mathrm{d}}

\newcommand{\DD}{{\rm{D}}}
\newcommand{\tmb}[1]{{\mbox{\tiny{#1}}}}
\newcommand{\NambuG}{Nambu-Got\={o}\;}
\newcommand{\chisqred}{\chi^2_\mathrm{red}}

\title{Studying Effective String Theory using deep generative models}
\author[a]{Michele Caselle}
\author*[a]{Elia Cellini}
\author[a]{Alessandro Nada}

\affiliation[a]{Dipartimento di Fisica,  Universit\'a degli Studi di Torino and INFN, Sezione di Torino, \\
  Via Pietro Giuria 1, I-10125 Turin, Italy}

\emailAdd{elia.cellini@unito.it}

\abstract{Effective String Theory (EST) offers a robust non-perturbative framework for describing confinement in Yang-Mills theory by treating the confining flux tube between a static quark-antiquark pair as a thin, vibrating string. While EST calculations are typically carried out using zeta-function regularization, certain problems—such as determining the flux tube width—are too complex to solve analytically. However, recent studies have demonstrated that EST can be explored numerically by employing deep learning techniques based on generative algorithms. In this work, we provide a brief introduction to EST and this novel numerical approach. Finally, we present results for the width of the \NambuG EST.}

\FullConference{The XVIth Quark Confinement and the Hadron Spectrum Conference (QCHSC24)\\
 19-24 August, 2024\\
Cairns Convention Centre, Cairns, Queensland, Australia\\}
\begin{document}
\maketitle

\section{Introduction}
The study of confinement is essential for advancing our understanding of the non-perturbative regime of Quantum Chromodynamics (QCD), particularly in explaining the absence of isolated color charges and the emergence of hadronic bound states. Among the various theoretical approaches to study confinement in pure Yang-Mills theories, such as the study of monopoles and center of vortex~\cite{Mickley:2024vkm, Crean:2024nro, Junior:2024urr,Mickley:2025mjj}, a powerful framework emerged in the past few years is represented by Effective String Theory (EST). In the latter, the chromoelectric flux tube joining couples of quark-antiquark is modeled as a vibrating string~\cite{Nambu:1974zg, Goto:1971ce, Luscher:1980ac, Luscher:1980fr, Polchinski:1991ax}. In particular, the Polyakov loops correlators at distance $R$ are associated with the full partition function of an EST action:
$$\langle P(0) P^{\dagger}(R)\rangle \sim \int \DD X \; e^{-S_\tmb{EST}[X]}.$$

The most straightforward choice for the EST action is the \NambuG string. While it exhibits anomalies at the quantum level, these disappear in the large-distance limit. Various analytical investigations of the \NambuG partition function, along with lattice computations of pure gauge theories, have demonstrated that EST serves as a highly predictive effective model~\cite{Aharony:2013ipa,Brandt:2016xsp,Caselle:2021eir,Caristo:2021tbk,Baffigo:2023rin,Caselle:2024zoh}. However, certain aspects—such as analyzing the flux tube width—pose challenges that analytical methods alone cannot resolve. A different numerical strategy for studying EST consists of discretizing the EST functional on the lattice and subsequently applying Markov Chain Monte Carlo (MCMC) techniques to sample from the associated probability distribution. Nevertheless, due to the strong non-linearity inherent in EST actions, conventional MCMC approaches experience critical slowing down when simulating complex EST regimes~\cite{Caselle:2023mvh}.

In order to address the numerical problems of lattice regularized EST, refs.~\cite{Caselle:2023mvh,Caselle:2024ent} presented a new numerical method based on Deep Learning (DL) sampling techniques. DL, together with quantum computing, provide a novel and promising route to overcome the problems of standard lattice simulations. The former, in particular through a class of algorithms called Normalizing Flows (NFs)~\cite{rezende2015variational}, have been proposed to tackle critical slowing down while the latter allows for real-time sign problem free simulations of lattice gauge theories through the Hamiltonian formalism~\cite{Banuls:2019bmf,Crippa:2024hso}. 

NFs are a class of deep generative models that can generate uncorrelated samples from a given target distribution~\cite{Albergo1}. Additionally, they provide a means to compute observables, including partition functions, which remain inaccessible to conventional MCMC methods~\cite{Nicoli:2019gun,Nicoli2021}.

Although flow-based samplers offer significant advantages, they encounter difficulties when scaling to state-of-the-art applications~\cite{DelDebbio,Abbott:2022zhs,Abbott:2023thq}. These challenges can be addressed by integrating Normalizing Flows (NFs) with non-equilibrium simulations rooted in out-of-equilibrium statistical mechanics~\cite{Caselle:2016wsw,CaselleSU3,Francesconi:2020fgi,Bulgarelli:2023ofi,Bulgarelli:2024onj,Bonanno:2024udh,Bonanno:2024fkn}, leading to the development of Stochastic Normalizing Flows (SNFs)~\cite{wu2020stochastic, Caselle:2022acb}. SNFs have shown remarkable scalability across both scalar and gauge theories on the lattice, positioning them among the most promising samplers in lattice field theory~\cite{Caselle:2024ent,Bulgarelli:2024yrz,Bulgarelli:2024brv}. Notably, in ref.~\cite{Caselle:2024ent}, SNFs successfully overcame the scaling limitations encountered in the first study of lattice EST~\cite{Caselle:2023mvh}.

In this proceeding, we first briefly review the concepts of the lattice \NambuG EST and NFs; afterward, we present numerical results for the width of the EST, a correlation function related to the density of the chromoelectric flux tube.

\section{Lattice \NambuG String}
For a $d=2+1$ target Yang-Mills theory, the corresponding \NambuG action, when regularized on a two-dimensional lattice using a “physical gauge,” takes the form~\cite{Caselle:2023mvh}:
\begin{equation}\label{eq:NG}
S_\tmb{NG}=\sigma \sum_{x \in \Lambda} \biggl(\sqrt{1+\bigl((\partial_{x^0}\phi(x))^2+(\partial_{x^0}\phi(x))^2\bigr)/\sigma}-1\biggr)
\end{equation}
where $\Lambda$ denotes a square lattice of dimensions $L\times R$ with index $x=(x^0,x^1)$ describing the worldsheet of the string, and a lattice spacing set to $a=1$. The field $\phi(x) \in \mathbb{R}$ represents the transverse degrees of freedom of the string, while $\sigma$ corresponds to the string tension, which serves as the coupling parameter of the theory. We impose periodic boundary conditions $\phi(x^0,x^1)=\phi(x^0+L,x^1)$ along the temporal direction of length $L$, and Dirichlet boundary conditions $\phi(x^0,0)=\phi(x^0,R)=0$ along the spatial direction of length $R$, representing the Polyakov loops $P$ and $P^{\dagger}$. With this setup, the physical temperature can be identified as the inverse of $L$.

The width of the regularized \NambuG string can be computed as:
\begin{equation}
 \sigma w^2(\sigma, L,R)=\langle \phi^2(x^0,R/2)\rangle_{x^0}    
\end{equation}
where the expectation value $\langle ... \rangle_{x^0}$ is evaluated also over the temporal extension $x^0$ (because of the translation invariance imposed by the periodic boundary conditions). The only analytical solution for this observable is a perturbative calculation up to the next-to-leading order~\cite{Gliozzi:2010jh, Gliozzi:2010zt, Gliozzi:2010zv}.
In the high-temperature regime $R \gg L$ the expected behaviour is
\begin{equation}
\label{eq:GPHT}
 \sigma w^2(\sigma, L, R) = \frac{1}{2\pi} \log \frac{L}{L_c} + \frac{R}{4L} + \frac{\pi}{24}\frac{R} {\sigma L^3}+\cdots
\end{equation}
where $L_c$ is a new scale which emerges from the regularization of the correlator defining the string width and represents the limit under which the EST prediction are not valid.

In the high-temperature limit it has been conjectured, using the Svetitsky-Yaffe mapping of the lattice gauge theory into a suitable two-dimensional spin model, that the terms multiplying the $R/4L$ factor can be resummed~\cite{Caselle:2010zs}:
$$\biggl(1+\frac{\pi}{6\sigma L^2}+\cdots\biggr) = \frac{1}{\sqrt{1 -  \frac{\pi}{3 \sigma L^2}}} = \frac{\sigma}{\sigma(L)}.$$
where:
\begin{equation}\label{eq:sigmaL}
    \sigma(L)=\sigma\sqrt{1-\frac{\pi}{3\sigma L^2}}
\end{equation}
encodes the dependence of the string tension on the physical temperature.
Thus, the behaviour of the linear term in $R$ of the \NambuG string width is expected to be:
\begin{equation}
\label{eq:w2NGHT}
    w^2(\sigma, L, R) = \frac{1}{\sigma(L)}\frac{R}{4L} + \cdots
\end{equation}

A reliable numerical test of the next-the-leading order and of the conjecture is the main goal of this contribution.

\section{Flow-based samplers}
In recent years, flow-based samplers have emerged as one of the most promising deep learning-based approach to overcome the problems of standard lattice simulations~\cite{Albergo1,Nicoli2021,Cranmer:2023xbe}. Normalizing Flows (NFs)~\cite{rezende2015variational}, are a class of deep generative models that allows for the exact evaluation of the density $q_\theta$, of the generated samples. The main idea behind NFs is to optimize the variational density $q_\theta$ to approximate a given target Boltzmann distribution $p\propto \exp{-S}$.

The training procedure in done by minimizing with respect to the parameters of the NFs $\theta$ the so-called \textit{reverse Kullback-Leibler divercence}:
\begin{equation}\label{eq:KLNF}
\DKL( q_\theta || p) = \int \dd \phi q_\theta \ln \frac{q_\theta(\phi)}{p(\phi)} \propto \int \dd \phi q_\theta \bigl(\ln q_\theta(\phi)+S(\phi)\bigr)
\end{equation}
Once the training is done, one can compute asymptotically unbiased estimator for the vacuum expectation values over the target $p$ using a re-weighting procedure generally called Importance Sampling (IS) in the machine learning field~\cite{Nicoli:2019gun,Nicoli2021}:
\begin{equation}
\begin{split}
\langle \mathcal{O} (\phi)\rangle_{\phi \sim p} &= \int \dd p(\phi) \phi \mathcal{O} (\phi)= \int \dd \phi q_\theta(\phi) \frac{p(\phi)}{q_\theta(\phi)} \mathcal{O}(\phi) \\
 & = \frac{\int\dd \phi q_\theta(\phi) \mathcal{O}(\phi)e^{-S(\phi)-\ln q_\theta(\phi)}}{\int \dd \phi q_\theta e^{-S(\phi)-\ln q_\theta(\phi)}}
\end{split}
\end{equation}

Among the various NF architectures, two remarkable examples in lattice field theory are Continuous NFs (CNFs)~\cite{chen:2018,Gerdes:2022eve} and Stochastic NFs (SNFs)~\cite{wu2020stochastic, Caselle:2022acb, Bulgarelli:2024brv}. The former generate the samples using Neural Ordinary Differential Equation~\cite{chen:2018} while the latter combine standard NFs with Non-Equilibrium Markov Chain Monte Carlo algorithms~\cite{Caselle:2016wsw,CaselleSU3,Francesconi:2020fgi,Bulgarelli:2023ofi,Bulgarelli:2024onj,Bonanno:2024udh,Bonanno:2024fkn}. An intriguing aspect of SNFs is that, in this setup, the IS described above generalizes to the Jarzynski's equality~\cite{Jarzynski1997}, a milestone in the field of non-equilibrium thermodynamics. 

In EST, CNFs have been used in the first work on this subject to provide a proof-of-concept for the feasibility of the application of flow-based samplers to EST lattice calculations~\cite{Caselle:2023mvh}. On the other hand, SNFs represent right now the most promising class of flow-based samplers and they have been used in ref.~\cite{Caselle:2024ent} to obtain state-of-the-art numerical results for the width and shape of the effective string. In the following, we show numerical results for the width obtained with both, CNFs and SNFs.

\section{Numerical Results}
\begin{figure}[ht]
  \centering
\includegraphics[scale=0.8,keepaspectratio=true]{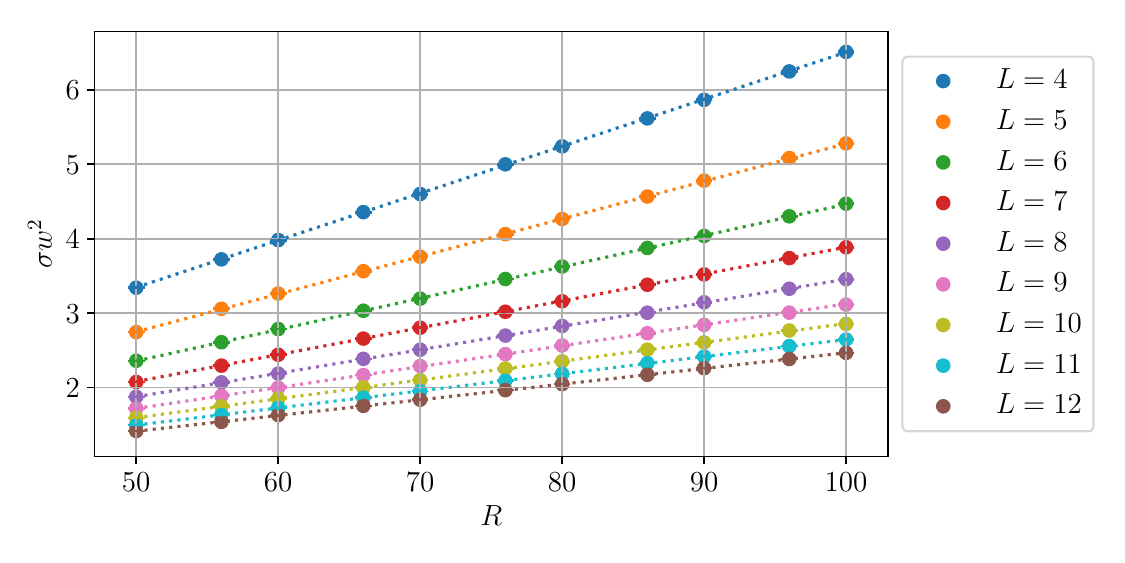}
  \caption{Plot of $\sigma w^2$ as a function of $R$, with $\sigma=100.0$ for various values of $L$. The dotted lines indicate the best fit, while the error bars are smaller than the data points.}
   \label{fig:HTthickness}
 \end{figure}

\begin{figure}
  \centering
\includegraphics[scale=0.8,keepaspectratio=true]{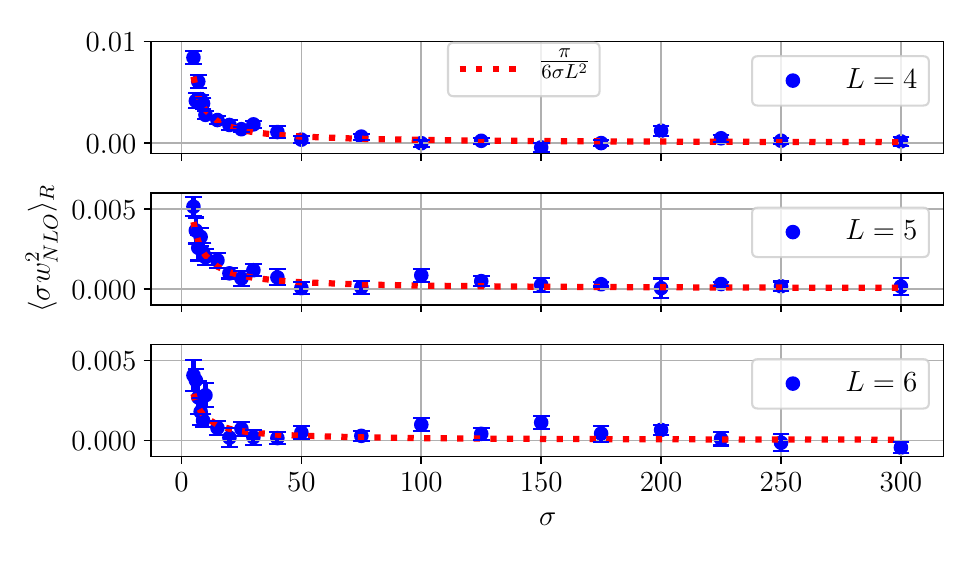}
  \caption{Plot of $\langle \sigma w^2_{NLO}\rangle_R$ as a function of $\sigma$ for three different values of $L$, compared with the expected analytical solution $\frac{\pi}{6 \sigma L^2}$.}
   \label{fig:HTPepe}
 \end{figure}

In this section, we study the width of the \NambuG effective string using CNFs and SNFs in the high-temperature regime: $R\gg L$. With the CNFs, we used same architecture, as mention in ref.~\cite{Caselle:2023mvh}. However, we found poor scaling in the string tension $\sigma$, and we could study only values $\sigma \geq 5$. Nevertheless, as we shall see below, in this $\sigma$-regime we could observe the linear broadening with $R$ of the width (see fig.~\ref{fig:HTthickness}) and, furthermore, we could observe the next-to-leading order correction computed in refs.~\cite{Gliozzi:2010jh, Gliozzi:2010zt, Gliozzi:2010zv}. We fit the results using the expression:
\begin{equation}
\label{eq:w2HT}
\begin{split}
    \sigma w^2(\sigma,L,R)=&\biggl(1+\frac{a^{(0)}}{\sigma}+\frac{a^{(1)}}{\sigma L^2}\biggr)  \biggl(b\frac{R}{L} + c + d\log (L) \biggr)    
\end{split}
\end{equation}
where the expected values for the leading term are $b=\frac{1}{4}$ and $d=\frac{1}{2\pi}=0.159155…$, while for the next-to-leading correction, we have $a^{(1)}_{HT}=\frac{\pi}{6}=0.523598…$. The remaining terms correspond to non-universal contributions that emerge in the lattice-regularized EST. Results are reported in table~\ref{tab:HTthickness}.  
\begin{table}
\centering
\begin{tabular}{|c c c c c c|} 
 \hline
 $a^{(0)}$ & $a^{(1)}$ & $b$ & $c$ & $d$ & $\chi^2/d.o.f.$\\ [0.5ex] 
 \hline\hline
 0.991(2) & 0.55(5) & 0.25007(4) & -0.032(1) & 0.1579(5) & 1.02 \\ 
 \hline
\end{tabular}
\caption{Coefficients obtained from fitting the string thickness in the HT regime, as given by eq.~(\ref{eq:w2HT}).}
\label{tab:HTthickness}
\end{table}
To isolate the next-to-leading contribution, we examined the following quantity:
\begin{equation}
\langle \sigma w^2_{NLO}\rangle_R(\sigma,L)=\langle\frac{\sigma w^2(\sigma,L,R)}{b\frac{R}{L}+c+d\log (L)}-1-\frac{a^{(0)}}{\sigma}\rangle_R,
\end{equation}
where we utilized the best-fit values for the coefficients $b$, $c$, $d$, and $a^{(0)}$, averaging over different values of $R$, as indicated by the notation $\langle … \rangle_R$. Specifically, we assume that the entire dependence of the width on $R$ is captured by the $R/L$ term in the denominator. Consequently, $\langle \sigma w^2_{NLO}\rangle_R$ is expected to follow the behavior $\frac{\pi}{6 \sigma L^2}$~\cite{Gliozzi:2010jh, Gliozzi:2010zt, Gliozzi:2010zv}. In fig.~\ref{fig:HTPepe}, we present this quantity as a function of $\sigma$ for three different values of $L$, alongside the expected behavior derived from the two-loop calculation~\cite{Gliozzi:2010jh, Gliozzi:2010zt, Gliozzi:2010zv}. The plots clearly illustrate the agreement with theoretical predictions.

\begin{figure}[h]
  \centering
  \centerline{\includegraphics[scale=0.8,keepaspectratio=true]{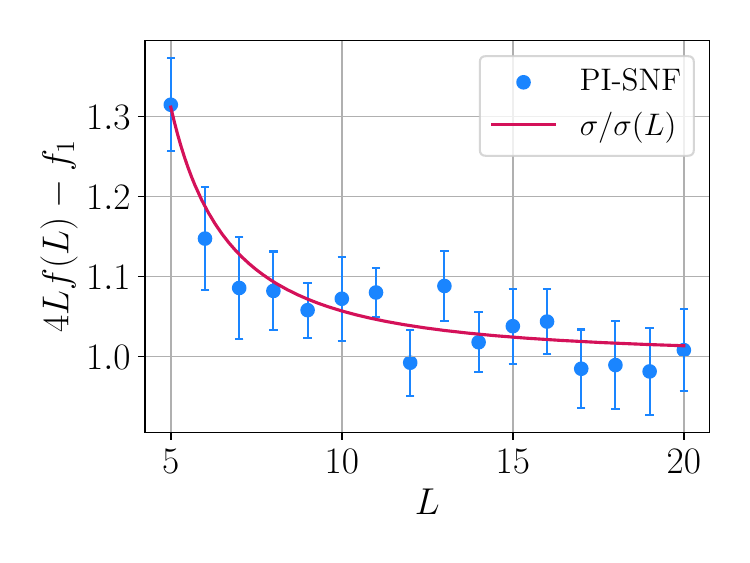}}
  \caption{Numerical results for the term $\sigma/\sigma(L)$ computed from the coefficient $f(l)$ by removing the non-universal term $f_1$ and multiplying by $4L$. The solid line represents the conjectured value from eq.~\eqref{eq:sigmaL}.}
  \label{fig:W2NG}
\end{figure}

On the other hand, using SNFs, we could overcome the scaling issues of the CNFs, and reach small values of $\sigma$. With this setup, we performed a study with $\sigma=0.1$, we fitted the $R$ dependence of our results with the following expression:
\begin{equation}
\label{eq:width1}
    \sigma w^2(L,R) = f(L) R + g(L);
\end{equation}
Then, we fit $L$ the coefficient $f(L)$ using
\begin{equation}
\label{eq:width2}
    f(L)= \frac{1}{4L} \left( \frac{1}{\sqrt{1 - \frac{f_0}{\sigma L^2}}} + f_1 \right)
\end{equation}
where $f_1$ is an undetermined bulk constant and $f_0$, should take the conjectured value $\pi/3$.
Results of the fit are reported in table~\ref{tab:widthNGf}. The results $f_0=1.01(9)$ remarkably agrees with the conjectured prediction $\pi/3=1.047...$. This finding can be clearly observed in fig.~\ref{fig:W2NG} as well, where the numerical prediction $4Lf(L)-f_1$, is compared with the conjecture behaviour $\sigma/\sigma(L)$.

\begin{table}[h]
    \centering
\begin{tabular}{|c c c|} 
 \hline
 $f_0$ & $f_1$ & $\chisqred$ \\ [0.5ex]
 \hline\hline
 $1.01(9)$ & $12.35(1)$ & $0.47$ \\
 \hline
\end{tabular}
    \caption{Results of the fit for eq.~\eqref{eq:width2}. }
    \label{tab:widthNGf}
\end{table}

\section{Conclusion}
In this contribution, using a rather unusual but totally new approach to EST calculations based on deep generative models, we show numerical results for the width of the \NambuG EST in the high-temperature regime, a challenging calculation for standard analytical method. Using a CNF architecture we provide a numerical test for the next-to-leading order of the width computed analytical in ref.~\cite{Gliozzi:2010jh, Gliozzi:2010zt, Gliozzi:2010zv}. Afterward, using a SNF, we reached small $\sigma$ and provided a numerical proof for the conjecture provided in ref.~\cite{Caselle:2010zs}.

The method used in this contribution could be extended to investigate other observables, as the shape of the flux tube~\cite{Caselle:2024ent,Verzichelli:2025cqc} as well as other models. Namely, an intriguing direction for future application is the extension of our approach to the $d=3+1$ case to study four-dimensional lattice gauge theories and the so-called axionic string~\cite{Dubovsky:2016cog, Athenodorou:2024loq}. Another interesting candidate for further numerical studies are the so-called higher-order corrections to the \NambuG theory, a more general class of EST that provide a better description of the confining flux tube close to the critical point~\cite{Caselle:2021eir,EliasMiro:2019kyf,Caselle:2024ent}.

\acknowledgments
We thank Andrea Bulgarelli, Marco Panero, Dario Panfalone and Lorenzo Verzichelli for several insightful discussions. We acknowledge support from the SFT Scientific Initiative of INFN. This work was partially supported by the Simons Foundation grant 994300 (Simons Collaboration on Confinement and QCD Strings) and by the European Union - Next Generation EU, Mission 4 Component 1, CUPD53D23002970006, under the Italian PRIN “Progetti di Ricerca di Rilevante Interesse Nazionale – Bando 2022” prot. 2022ZTPK4E.

\bibliographystyle{JHEP}
\bibliography{biblio}

\end{document}